\documentclass[prl,aps,twocolumn,showpacs]{revtex4}
\usepackage{epsfig}
\usepackage{bm}

\begin{document}

\title{Chaos from turbulence: stochastic-chaotic equilibrium in turbulent convection at high Rayleigh numbers}

\author{\small  A. Bershadskii}
\affiliation{\small {ICAR, P.O.B. 31155, Jerusalem 91000, Israel; \\
ICTP, Strada Costiera 11, I-34100 Trieste, Italy}}

\begin{abstract}
It is shown that correlation function of the mean wind velocity generated by a turbulent 
thermal convection (Rayleigh number $Ra \sim 10^{11}$) exhibits exponential decay with 
a very long correlation time, while corresponding largest Lyapunov exponent is 
certainly positive. These results together with the reconstructed phase portrait 
indicate presence of chaotic component in the examined mean wind. Telegraph approximation 
is also used to study relative contribution of the chaotic and stochastic components to 
the mean wind fluctuations and an equilibrium between these components has been studied in detail.

\end{abstract}


\maketitle

{\it Introduction}.  Turbulent thermal convection can generate large-scale (coherent) circulations, 
also known as mean winds. In recent decade a vigorous investigation of statistical properties of 
the thermal mean winds has been launched \cite{kad},\cite{sbn} (see for a recent review \cite{agl}). 
The winds' dynamics turned out to be very complex and many surprising features 
were discovered in laboratory experiments and numerical simulations. 
Simple stochastic models \cite{sbn},\cite{ben} were replaced by more sophisticated 
stochastic models \cite{agl},\cite{bv},\cite{ba} which also addressed significant three-dimensional nature 
of the phenomenon. In these models, interaction between the large-scale wind and the small-scale 
turbulence provides a phenomenological stochastic driving term. 
Also some deterministic models with chaotic solutions were recently suggested \cite{fon},\cite{res}, 
and the idea that a large-scale instability in the developed turbulent convection 
(caused by a redistribution of the turbulent heat flux) can be an origin of the large-scale 
coherent structures received certain experimental support (see Ref. \cite{bgu} and references therein).   

\begin{figure} \vspace{-0.5cm}\centering
\epsfig{width=.45\textwidth,file=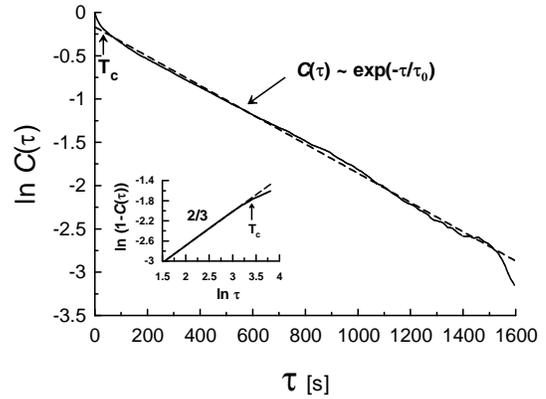} \vspace{-4.5cm}
\caption{Autocorrelation function of a mean wind velocity for a 
Rayleigh-Benard convection laboratory experiment \cite{sbn}. The dashed 
straight line indicates the exponential decay Eq. (1). The insert shows 
a small-time-scales part of the correlation function defect in ln-ln scales. The straight 
line indicates the Kolmogorov's '2/3' power law for structure function.}
\end{figure}
\begin{figure} \vspace{-1cm}\centering
\epsfig{width=.45\textwidth,file=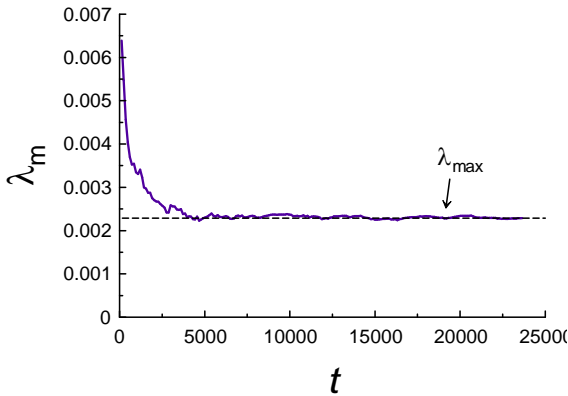} \vspace{-4.5cm}
\caption{The pertaining average maximal Lyapunov exponent 
at the pertaining time, calculated for the same data 
as those used for calculation of the correlation function (Fig. 1). 
The dashed straight line indicates convergence to a positive value.}

\end{figure}
\begin{figure} \vspace{-0.5cm}\centering
\epsfig{width=.45\textwidth,file=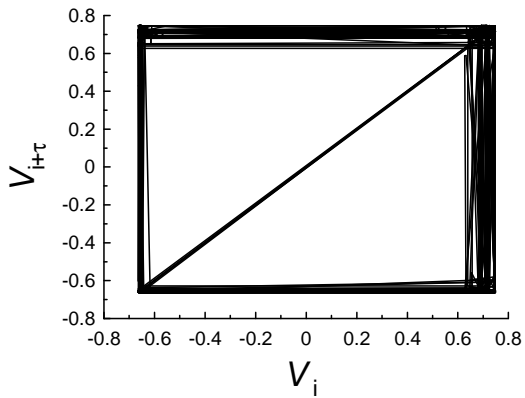} \vspace{-4cm}
\caption{Phase portrait reconstructed from noise-reduced time series.}
\end{figure}

It is a difficult problem to distinguish between stochastic and chaotic processes, especially 
having a developed turbulence as a small-scale background. Observation of the Lorentzian spectra 
for the mean winds (reported in the Ref. \cite{ba}) makes this problem even more difficult. 
It is shown recently (see, for instance, Ref. \cite{anis}) that many properties of the 
chaotic processes with exponentially decaying correlation function (i.e. with the Lorentzian spectra) 
can be reproduced by some classical models of stochastic processes (and vise versa). Therefore, 
for such chaotic systems stochastic phenomenological models can be rather useful (and relevant). 
However, it is necessary to reveal an underlying chaotic (deterministic) nature of the phenomenon 
in order to understand real cause of the long-term correlations in such systems.  \\

{\it Chaotic mean wind}. Figure 1 shows a correlation function of a mean wind velocity, 
$v(t)$, measured in a typical thermal convection 
laboratory experiment in a closed cylindrical container of aspect ratio 1 
for Rayleigh number $Ra \sim 10^{11}$ (see Ref. \cite{sbn},\cite{niem} for 
description of the experiment and of the other properties of the mean wind). The dashed straight 
line is drawn in the figure in order to indicate (in the semi-log scales) the exponential 
decay
$$
C(\tau) \sim \exp -(\tau/\tau_0)  \eqno{(1)}
$$
The correlation time $\tau_0 \simeq 600$s is very large in comparison with the mean wind circulation period 
$T_c \simeq 30$s. The circulation period (turnover time) provides a time scale relevant to the 
phenomenon physics. The correlation time normalized by this time scale $\tau_0/T_c \sim 20$. 

The insert in Fig. 1 shows a small-times part of the correlation function defect. 
The ln-ln scales have been used in this figure in order to show a power law (the dashed straight line) 
for structure function: $\langle (v(t+\tau)- v(t))^2 \rangle$ for $\tau < T_c$:
$$
1- C(\tau ) \propto  \langle (v(t+\tau)- v(t))^2 \rangle \propto \tau^{2/3}  \eqno{(2)}
$$  
This power law: '2/3', for structure function (by virtue of the Taylor hypothesis 
transforming the time scaling into the space one \cite{my}) 
is known for fully developed turbulence as Kolmogorov's power law (cf Refs. \cite{bgu},\cite{my}). 

In order to determine the presence of a deterministic chaos in the time series corresponding to 
the velocity of the mean wind for the time scales $T_c < t < \tau_0$, we calculated the largest Lyapunov exponent 
: $\lambda_{max}$. A strong indicator for the presence of chaos in the examined time series is condition 
$\lambda_{max} >0$. If this is the case, then we have so-called exponential instability. Namely, 
two arbitrary close trajectories of the system will diverge apart exponentially, that is 
the hallmark of chaos. At present time there is no theory relating $\lambda_{max}$ 
to $\tau_0^{-1}$ for chaotic systems. There is only a tentative suggestion that their values should be of the same order. 
To calculate $\lambda_{max}$ we used a direct algorithm developed by Wolf et al \cite{w}. Figure 2 shows 
the pertaining average maximal Lyapunov exponent at the pertaining time, calculated for the same data 
as those used for calculation of the correlation function (Fig. 1). The largest Lyapunov exponent converges very 
well to a positive value $\lambda_{max} \simeq 0.0023s^{-1}$. 

The ambivalent nature of the turbulence-induced coherent dynamics one can also see in Figure 3. 
This figure shows a phase portrait reconstructed from the noise-reduced time series for the mean wind velocity. 

Thus, we can conclude that the thermal wind under study exhibit chaotic features 
and the long-term exponential correlation (Fig. 1) is presumably related to this chaotic behavior.\\

{\it Telegraph approximation}. It is shown in Ref. \cite{jsp} that simple telegraph 
approximation of stochastic signals can reproduce main statistical properties 
of these signals. The telegraph approximation of signal $v(t)$ can be constructed as
following:
$$
u(t) = v (t)/|v (t)| \eqno{(3)}
$$
From the definition the telegraph approximation can take only two values: 1 and -1. 
Figure 4 shows a comparison between correlation function of the full signal for the 
mean wind velocity $v(t)$ (cf Fig. 1) and correlation function of its telegraph approximation 
$u(t)$. One can see very good correspondence between these correlation functions. 
Therefore, certain main statistical properties of the full signal can be studied using 
the telegraph approximation in this case as well. 

\begin{figure} \vspace{-0.5cm}\centering
\epsfig{width=.45\textwidth,file=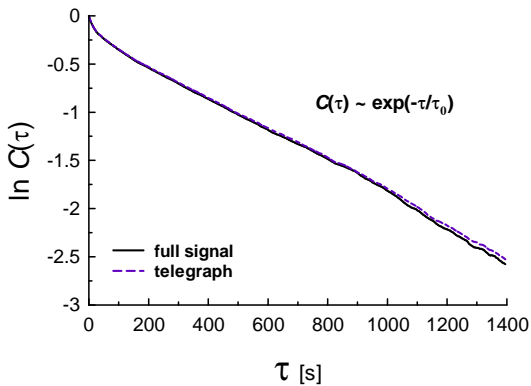} \vspace{-5cm}
\caption{Autocorrelation functions of a mean wind velocity for a 
Rayleigh-Benard convection laboratory experiment \cite{sbn} (solid curve) and 
of its telegraph approximation (dashed curve).}
\end{figure}
\begin{figure} \vspace{-0.5cm}\centering
\epsfig{width=.45\textwidth,file=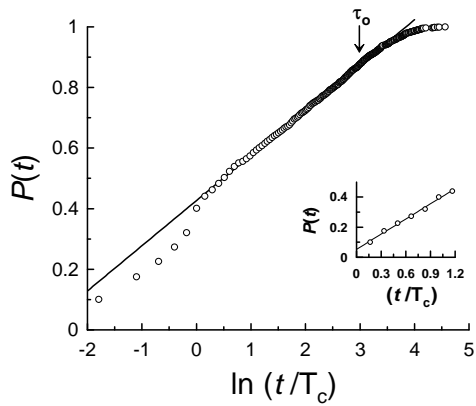} \vspace{-4.5cm}
\caption{Cumulative probability $P(t)$ (Eq. (5)) vs. $\ln (t/T_c) $. The straight line indicates correspondence to Eq. (6). The insert shows a linear dependence of $P(t)$ on $t/T_c$ for $t \leq T_c$.}
\end{figure}
 
Very significant characteristic 
of the telegraph signal is duration, $\tau $, of the continuous intervals (boxes) where the
signal takes value 1 or value -1. Probability density function of
the duration (or life) times $\tau$: $~~p(\tau)$, for the telegraph
approximation of the mean wind velocity signal was studied in detail in Ref. \cite{sbn}. 
It has a peak at $\tau \simeq T_c$. In the the range $T_c < \tau < \tau_0$ 
the probability density exhibits a power law
$$
p(\tau) \simeq c~ \tau^{\alpha}  \eqno{(4)}
$$
with the value of the exponent $\alpha$ close to $-1$. For $\tau > \tau_0$ the probability density 
$p(\tau)$ decays exponentially, as it should be for the random telegraph signal. Figure 5 shows 
cumulative probability
$$
P(t) = \int_0^t p(\tau) d\tau  \eqno{(5)}
$$
In the semi-logarithmical scales the straight line indicates just the power law for $p(\tau)$ (Eq. (4) with 
$\alpha \simeq -1$): 
$$
P(t) \approx P(T_c) + c \int_{T_c}^t \tau^{-1} d\tau   \approx P (T_c) + c~ \ln ( t/T_c ) \eqno{(6)}   
$$ 
where $P (T_c)$ is a constant part provided by the turbulent component to the cumulative probability 
$P(t)$ for $\tau_0 > t > T_c$. On the whole,
$$
P(\infty) \simeq \Delta P + c~ \ln ( \tau_0/T_c )  \eqno{(8)}
$$
where
$$
\Delta P = P(T_c) + \int_{\tau_0}^{\infty} p(\tau)~ d\tau  \eqno{(9)}
$$
i.e. $\Delta P$ is contribution of the {\it stochastic} components: turbulence and the large-scale 
random telegraph signal, to the total cumulative probability. 
In our case there is a probabilistic equipartition of the stochastic and chaotic components 
to the total cumulative probability $P(\infty)=1$: 
$~~\Delta P \simeq 0.5 \pm 0.04$. If this probabilistic equipartition is universal 
and the constant $c$ in the power law Eq. (4) is also universal (at least asymptotically), then 
$$
\ln ( \tau_0/T_c ) = \frac{1- \Delta P}{c} \approx 0.5/c \eqno{(10)}
$$
is universal as well. From the present data we can estimate: $c \approx 0.16 \pm 0.01$, and 
$\ln ( \tau_0/T_c ) \approx 3$ (see also below). \\

{\it Stochastic-chaotic equilibrium}. This kind of probabilistic equipartition 
is usually related to statistical restoration of a symmetry. For instance, the spontaneous appearing 
of the mean wind results in a spontaneous breaking of mirror (or parity - P) symmetry. 
Statistical restoration of the parity (P-symmetry) means probabilistic equipartition 
of the $u(t)=1$ and $u(t)=-1$ events (Eq. (3)). This probabilistic equipartition indeed takes place 
in present case and it is clear evidence of the statistical restoration of the P-symmetry (the directions 
of the mean wind rotation: clockwise and anticlockwise, are statistically equivalent). 
While the P-symmetry arises from space inversion, the T-symmetry 
arises from time reversal (reversibility). In the present case both P- and 
T-transformations result in the same mean wind switching: $u \rightarrow -u$. 
It can be just statistical restoration of the combined PT-symmetry 
(statistical invariance under joint action of parity and time reversal) that results in the equal 
probability of the reversible (chaotic) and irreversible (random) switchings. 
The statistical restoration of the symmetry in a mixed stochastic-chaotic motion indicates an equilibrium 
reached between the stochastic and chaotic components (SC-equilibrium).

  For the closed orbits in the chaotic systems the Bowen's theorem \cite{suz} 
is an analogue of ergodic theorem. This theorem provides a basis for the equality of the average 
over the phase space and the average over large-period orbits. Let us now, in the terms of the 
Bowen's theorem (cf. also \cite{az}), consider a phase volume subregion 
$\Delta \Omega$: $~~ \Omega_{c} < \Delta \Omega \ll \Omega$ 
(where $\Omega_{c}$ is phase volume corresponding to the upper turbulent scale $T_c$) and 
all the periodic trajectories passing through it. Different closed orbits can be distinguished by their period $T$. 
Their distribution does not depend significantly on location and boundaries of the subregion $\Delta \Omega$. 
Moreover, if we consider the periodic orbits with periods in the interval $(T,~ T+ t)$ passing through the region, then we can use an estimate
$$
\frac{t }{T_c} \approx \frac{\Delta \Omega}{\Omega_{c}} \eqno{(11)}
$$  
(the coarse graining is defined by the partition of phase space by the cells $\Omega_{c}$). 
Let us consider an entropy corresponding to the subregion $\Delta \Omega$ (cf. Ref. \cite{az}) 
and use estimate Eq. (11)
$$
S = \beta~ \ln \frac{\Delta \Omega}{\Omega_{c}} \approx \beta ~\ln \frac{t }{T_c} \eqno{(12)}
$$
where $\beta$ is a constant (see below, and cf. also Ref. \cite{gua} for quantum chaos). 
The logarithmic growth of entropy can imply certain self-similarity of joint probabilistic properties of the time series $v(t) \leftrightarrow u(t)$ 
(see Eq. (3)). Indeed, let us consider probability distribution function, $p(v,t)$, for the signal $v$ 
at the impulses (boxes) of length $\tau \leq t$ of its telegraph approximation $u$. For the stationary signal the self-similarity has a standard form \cite{bar}
$$
p(v,t)=\frac{1}{t^{\beta}}~f\left(\frac{v}{t^{\beta}}\right) \eqno{(13)}
$$
where $f$ is certain function of the argument $v/t^{\beta}$ (the statistical restoration of the parity 
has been taken into account here). Let us consider an entropy
$$
S(t) = -\int dv ~p(v,t)~\ln p(v,t)  \eqno{(14)}
$$
Changing the integration variable: $v \rightarrow v/t^{\beta}$ and using equilibrium condition: 
$S(T_c)=0$, one obtains from Eqs. (13) and (14): 
$$
S(t) = \beta ~\ln (t/T_c )  \eqno{(15)}
$$
(cf Eq. (12)).

The insert in Fig. 5 shows a linear dependence of $P(t)$ on $t/T_c$ for the turbulent regime that can be 
related to an analytic dependence of $P(t)$ on $t/T_c$ for $0 < t/T_c \leq 1$ (the linear dependence represents 
the first two terms approximation of the Taylor expansion; the small 'jump' at the point $t=0$ 
is presumably related to transition from laminar to turbulent motion). 
In a vicinity of the SC-equilibrium $P(t)$ is an analytic function of the entropy 
$S(t)=\beta \ln (t/T_c)$. Applying again the Taylor expansion we obtain
$$
P(t) = P(T_c) + c_o S(t) +... \approx P(T_c) + c_o \beta~\ln (t/T_c) \eqno{(16)}
$$
where $c_o = (\partial P/ \partial S)|_{S=0}$ (cf. Fig. 5 and Refs. 
\cite{az},\cite{hoo} for the transformation $t \rightarrow \ln t$). \\

{\it Discussion}.  It is possible that the above consideration is also applicable for 
the large-scale coherent structures observed in other turbulent flows 
(in turbulent boundary layers, for instance \cite{s}). 
If, for instance, at the SC-equilibrium the exponent $\beta$ is a global invariant of motion and 
its value is the same for $t < T_c$ and for $t > T_c$, then, taking into account that the dimensionless 
entropy $S(\tau_0) =1$ one obtains from Eq. (15): $~ \ln (\tau_0/T_c) =1/\beta~$. 
For the Brownian processes $\beta =1/2$, whereas for the Kolmogorov's scaling: Eq. (2), one has 
$\beta = 1/3$. For Kolmogorov's turbulence as a background it results in estimate: $~ \ln (\tau_0/T_c) =3~$ (cf. above). If the background turbulence produces scaling different from the Kolmogorov's one, 
then value of the $\beta$-exponent can be different 
from $1/3$. For the Kraichnan's scaling, for instance \cite{clv}, $\beta =1/4$ and, consequently, 
$ \ln (\tau_0/T_c) =4$ (i.e. in the case of the Kraichnan's background turbulence the 
correlations related to the large-scale coherent structures are even more long-term than those 
generated by the Kolmogorov's background turbulence). \\

The author is grateful to J.J. Niemela, to K.R. Sreenivasan for 
sharing their data and discussions.

\end{document}